\documentclass[ALICE,manyauthors]{cernphprep}
\usepackage{xcolor,pict2e,curve2e,lineno}

\newcommand{\Fi}[1]   {Fig.~\ref{#1}}

\newcommand{\Ta}[1]   {Table~\ref{#1}}

\newcommand{\gevc}    {\mbox{GeV$/c$}}
\newcommand{\gevcc}   {\mbox{GeV$/c^2$}}
\newcommand{\mevc}    {\mbox{MeV$/c$}}

\newcommand{\mum}     {\mbox{$\mu$m}}
\newcommand{\rb}[1]   {\mbox{\textrm{\scriptsize #1}}}
\newcommand{\rbt}[1]  {\mbox{\textrm{\tiny #1}}}

\newcommand{\pizero}  {\ensuremath{\pi^{0}}}

\newcommand{\jpsi}    {\ensuremath{\textrm{J}/\psi}}

\newcommand{\mpmm}    {\ensuremath{\mu^{+} \mu^{-}}}
\newcommand{\epem}    {\ensuremath{\textrm{e}^{+} \textrm{e}^{-}}}

\newcommand{\elp}     {\ensuremath{\textrm{e}^{+}}}
\newcommand{\elm}     {\ensuremath{\textrm{e}^{-}}}
\newcommand{\qqbar}   {\ensuremath{\textrm{q}\bar{\textrm{q}}}}
\newcommand{\ccbar}   {\ensuremath{\textrm{c}\bar{\textrm{c}}}}
\newcommand{\sqrts}   {\ensuremath{\sqrt{s}}}
\newcommand{\sqrtsnn} {\ensuremath{\sqrt{s_{_{\rbt{NN}}}}}}
\newcommand{\pt}      {\ensuremath{p_{\rb{t}}}}
\newcommand{\ptavg}   {\ensuremath{\langle p_{\rb{t}} \rangle}}

\newcommand{\minv}    {\ensuremath{m_{\rb{inv}}}}

\newcommand{\dedx}    {\ensuremath{\textrm{d}E/\textrm{d}x}}

\newcommand{\dndyjp}  {\ensuremath{\textrm{d}N_{\rb{J/}\psi}/\textrm{d}y}}

\newcommand{\dndetach}{\ensuremath{\textrm{d}N_{\rb{ch}}/\textrm{d}\eta}}

\newcommand{\chisq}   {\ensuremath{\chi^{2}}}
\newcommand{\vtxz}    {\ensuremath{z_{\rb{vtx}}}}
\begin{document}
\begin{titlepage}
\PHnumber{2012-021}
\PHdate{\today}
\title{\jpsi\ Production as a Function of Charged Particle
       Multiplicity in pp Collisions at \sqrts~= 7~TeV}
\ShortTitle{\jpsi\ Production as a Function of Charged Particle
  Multiplicity}
\Collaboration{The ALICE Collaboration
\thanks{See Appendix~\ref{app:collab} for the list of collaboration
  members}}
\ShortAuthor{The ALICE Collaboration} 
\vfill

%
\begin{abstract}
The ALICE collaboration reports the measurement of the inclusive
\jpsi\ yield as a function of charged particle pseudorapidity density
\dndetach\ in pp collisions at \sqrts~= 7~TeV at the LHC.  \jpsi\
particles are detected for $\pt > 0$, in the rapidity interval 
$|y| < 0.9$ via decay into \epem, and in the interval $2.5 < y < 4.0$
via decay into \mpmm\ pairs.  An approximately linear increase of the
\jpsi\ yields normalized to their event average
$(\dndyjp)/\langle\dndyjp\rangle$ with
$(\dndetach)/\langle\dndetach\rangle$ is observed in both rapidity
ranges, where \dndetach\ is measured within $|\eta| < 1$ and $\pt >
0$.  In the highest multiplicity interval with
$\langle\dndetach(\textrm{bin})\rangle = 24.1$, corresponding to
four times the minimum bias multiplicity density, an enhancement
relative to the minimum bias \jpsi\ yield by a factor of about 5 at
$2.5 < y < 4$ (8 at $|y| < 0.9$) is observed.
\end{abstract}
\vspace{2cm}
\Submitted{(Submitted to PLB)}
\end{titlepage}

%

Understanding the production mechanism of quarkonium states in
hadronic collisions is still a challenge due to its sensitivity to
perturbative and non-perturbative aspects of Quantum Chromodynamics
(QCD).  While the primary production of heavy quark anti-quark
(\qqbar) pairs is generally treated as a hard process which can be
calculated within perturbative QCD, the subsequent formation of a bound
colorless \qqbar~pair is inherently non-perturbative and difficult to
treat.  The models developed to describe quarkonium production in high
energy hadron collisions consequently follow various approaches,
mainly differing in the relative contribution of the intermediate
color singlet and color octet \qqbar~states
\cite{Brambilla:2010cs,Lansberg:2008gk}.  Recent theoretical work
tries to describe consistently
\cite{Ma:2010yw,Butenschoen:2010rq,Butenschoen:2012px} the measured
production cross section and polarization, in particular in light of
recent measurements at the LHC
\cite{Aad:2011sp,Aaij:2011jh,Aamodt:2011gj,Abelev:2011md,Khachatryan:2010yr,Chatrchyan:2011kc}.

It is also important to consider that a high energy proton-proton
collision can have a substantial contribution from Multi-Parton
Interactions (MPI) \cite{Sjostrand:1987su,Bartalini:2010su}.  In this
case several interactions on the parton level can occur in a single pp
collision, which can introduce a dependence of particle production on
the total event multiplicity
\cite{Acosta:2004wqa,Khachatryan:2010pv,Aad:2010fh}.  If MPI were
mainly affecting processes involving only light quarks and gluons, as
implemented e.g. in PYTHIA~6.4, processes like \jpsi\ and open heavy
flavour production should not be influenced and their rates are
expected to be independent of the overall event multiplicity.
However, at the high center-of-mass energies reached at the LHC, there
might be a substantial contribution of MPI on a harder scale which can
also induce a correlation between the yield of quarkonia and the total
charged particle multiplicity \cite{Porteboeuf:2010dw}.  An early
study that relates open charm production and underlying event
properties was performed by the NA27 experiment for pp collisions at
\sqrts~= 27~GeV, with the result that charged particle multiplicity
distributions in events with open charm production have a mean that is
higher by $\sim 20$\% than the ones without \cite{Aguilar:1988aa}.  

In \cite{Frankfurt:2008vi,Strikman:2011zz} it has been argued that,
due to the spatial distribution of partons in the transverse plane (as
described in generalized parton distributions), the density of partons
in pp collisions will be strongly impact parameter dependent.
Therefore, the probability for MPI to occur will increase towards
smaller impact parameters.  This effect might be further enhanced by
quantum-mechanical fluctuations of the small Bj{\o}rken-$x$ gluon
densities.

The charged particle multiplicities measured in high-multiplicity pp
collisions at LHC energies reach values that are of the same order as
those measured in heavy-ion collisions at lower energies (e.g. they
are well above the ones observed at RHIC for peripheral Cu--Cu
collisions at \sqrtsnn~= 200~GeV \cite{Alver:2010ck}).  Therefore, it
is a valid question whether pp collisions also exhibit any kind of
collective behaviour as seen in these heavy-ion collisions.  An
indication for this might be the observation of long range, near-side
angular correlations (ridge) in pp collisions at \sqrts~= 0.9, 2.36
and 7~TeV with charged particle multiplicities above four times the
mean multiplicity \cite{Li:2011mp,Khachatryan:2010gv}.  Since
quarkonium yields in heavy-ion reactions are expected to be modified
relative to minimum bias pp collisions
\cite{Matsui:1986,Andronic:2003zv,Miao:2010tk}, one might ask whether
their production rates in high-multiplicity pp collisions are already
exhibiting any effect like \jpsi\ suppression.

In this Letter, we report the first measurement of relative \jpsi\
production yields $(\dndyjp)/\langle\dndyjp\rangle$ at mid-rapidity
($|y| < 0.9$) and at forward rapidity ($2.5 < y < 4$) as a function of
the relative charged particle multiplicity density
$(\dndetach)/\langle\dndetach\rangle$ as determined in $|\eta| < 1$
for pp collisions at \sqrts~= 7~TeV at the LHC. 

The data discussed here are measured in two complementary parts of the
experimental setup of ALICE \cite{Aamodt:2008zz}: the central barrel
($|\eta| < 0.9$) for the \jpsi\ detection in the di-electron channel
and the muon spectrometer ($-4 < \eta < -2.5$) \footnote{In the
ALICE reference frame the muon spectrometer is located at negative $z$
positions and thus negative \mbox{(pseudo-)rapidities}.  Since pp
collisions are symmetric relative to $y = 0$, we have dropped the
minus sign when rapidities are quoted.} for $\jpsi \rightarrow \mpmm$
measurements.  

The central barrel provides momentum measurement for charged
particles with $\pt > 100$~\mevc\ and particle identification up to
$\pt \approx 10$~\gevc.  Its detectors are all located inside a large
solenoidal magnet with a field strength of 0.5~T.  Used in this
analysis are the Inner Tracking System (ITS) and the Time Projection
Chamber (TPC).  The ITS \cite{Aamodt:2010ys} consists of six layers of
silicon detectors surrounding the beam pipe at radial positions
between 3.9~cm and 43.0~cm.  Silicon Pixel Detectors (SPD) are used
for its innermost two layers and allow a precise reconstruction of the
interaction vertex.  The TPC \cite{Alme:2010ke} is a large cylindrical
drift volume covering the range along the beam axis relative to the
Interaction Point (IP) between $-250 < z < 250$~cm and extending in
the radial direction from 85~cm to 247~cm.  It is the main tracking
device in the central barrel and is also used for particle
identification via a measurement of the specific ionization (\dedx) in
the detector gas with a resolution of about 5\% \cite{Aamodt:2008zz}.

The muon spectrometer consists of a frontal absorber followed by a
3~T$\cdot$m dipole magnet, coupled to tracking and triggering
detectors.  Muons are filtered by the 10 interaction length
($\lambda_{\rb{I}}$) thick absorber placed between 0.9~m and 5.0~m
from the nominal position of the IP along the beam axis.  Muon
tracking is performed by five tracking stations, positioned between
5.2~m and 14.4~m from the IP, each consisting of two planes of cathode
pad chambers.  The muon triggering system consists of two stations
positioned at 16.1~m and 17.1~m from the IP, each equipped with two
planes of resistive plate chambers.  It is located downstream of a
1.2~m thick iron wall (7.2 $\lambda_{\rb{I}}$) which absorbs hadrons
penetrating the frontal absorber, secondary hadrons escaping the
absorber material, and low-momentum muons ($p < 4$~\gevc).  A conical
absorber surrounding the beam pipe provides protection against
secondary particles throughout the full length of the muon
spectrometer.

Two VZERO detectors are used for triggering on inelastic pp
interactions and for the rejection of beam-gas events.  They consist
of scintillator arrays and are positioned at $z = -90$~cm and $z =
+340$~cm, covering the pseudorapidity ranges $-3.7 < \eta < -1.7$ and
$2.8 < \eta < 5.1$.  The minimum bias (MB) pp trigger uses the
information of the VZERO detectors and the SPD.  It is defined as the
logical OR between two conditions: (i) a signal in at least one of the
two VZERO detectors has been measured; (ii) at least one readout chip
in the SPD fires.  It has to be in coincidence with the arrival of
proton bunches from both sides of the interaction region.  The
efficiency of the MB trigger to record inelastic collisions was
evaluated by Monte Carlo studies and is 86.4\% \cite{Oyama:2011aa}.
For the di-muon analysis, a more restrictive trigger is used
($\mu$-MB).  It requires the detection of at least one muon above a
threshold of $\pt^{\rb{trig}} >$~0.5~\gevc\ in the muon trigger
chambers in addition to the MB trigger requirement.  

The results presented in this Letter are obtained by analyzing pp
collisions at \sqrts~=~7~TeV recorded in 2010.  For the \jpsi\
measurement in the di-electron (di-muon) channel a sample of $3.0
\times 10^{8}$ minimum bias events ($6.75 \times 10^{6}$ $\mu$-MB
triggered events) is analysed, corresponding to an integrated
luminosity of 4.5~nb$^{-1}$ (7.7~nb$^{-1}$).  The di-electron sample
is divided into four separate data sets with slightly different
running conditions.  The relative normalization between the number of
$\mu$-MB and minimum bias triggers needed to extract the integrated
luminosity in the di-muon case is calculated using the ratio of the
number of corresponding single muons with $\pt > 1$~\gevc.  The
luminosity at the ALICE interaction point was kept between 0.6 and
1.2~$\cdot 10^{29}$~cm$^{-2}$~s$^{-1}$ for all the data used in this
analysis.  This ensures a collision pile-up rate of 4\%, or lower,
in each bunch crossing.  In the case of the di-muon analysis the
interaction vertex is reconstructed using tracklets which are defined
as combinations of two hits in the SPD layers of the ITS, one hit in
the inner layer and one in the outer.  Since for MB trigger used in
the di-electron analysis the full information of the central barrel
detectors is available ($\mu$-MB triggered events only include SPD
information), tracks measured with ITS and TPC are used in this case
to locate the interaction vertex.  This results in a resolution in
$z$~direction of $\sigma_{\rb{z}} \approx
600/N_{\rb{trk}}^{0.7}$~\mum, where $N_{\rb{trk}}$ is the multiplicity
measured via SPD tracklets.  For the vertices reconstructed using SPD
tracklets only, this resolution is worse by 35\% for high ($N_{\rb{trk}}
= 40$) and 50\% for low ($N_{\rb{trk}} = 10$) multiplicities.  Events
that do not have an interaction vertex within $|\vtxz| < 10$~cm are
rejected, where \vtxz\ is the reconstructed $z$~position of the
vertex.  The rms of the vertex distributions along $z$ is for all
running conditions below 6.6~cm.

Pile-up events are identified by the presence of two interaction
vertices reconstructed with the SPD.  They are rejected if the
distance along the beam axis between the two vertices is larger than
0.8~cm, and if both vertices have at least three associated tracklets.
This removes 48\% of the pile-up events.  In the remaining cases two
events can be merged into a single one, thus yielding a biased
multiplicity estimation.  A simulation assuming a Gaussian
distribution for the vertex $z$ position results in a probability for
the occurrence of two vertices closer than 0.8~cm of 7\%.  Combined
with the pile-up rate of 4\%, this gives an overall probability that
two piled-up events are merged into a single event of $\approx
0.3$\%, which is a negligible contribution in the multiplicity ranges
considered here.

%
\begin{figure}[t]
\includegraphics[width=\linewidth]{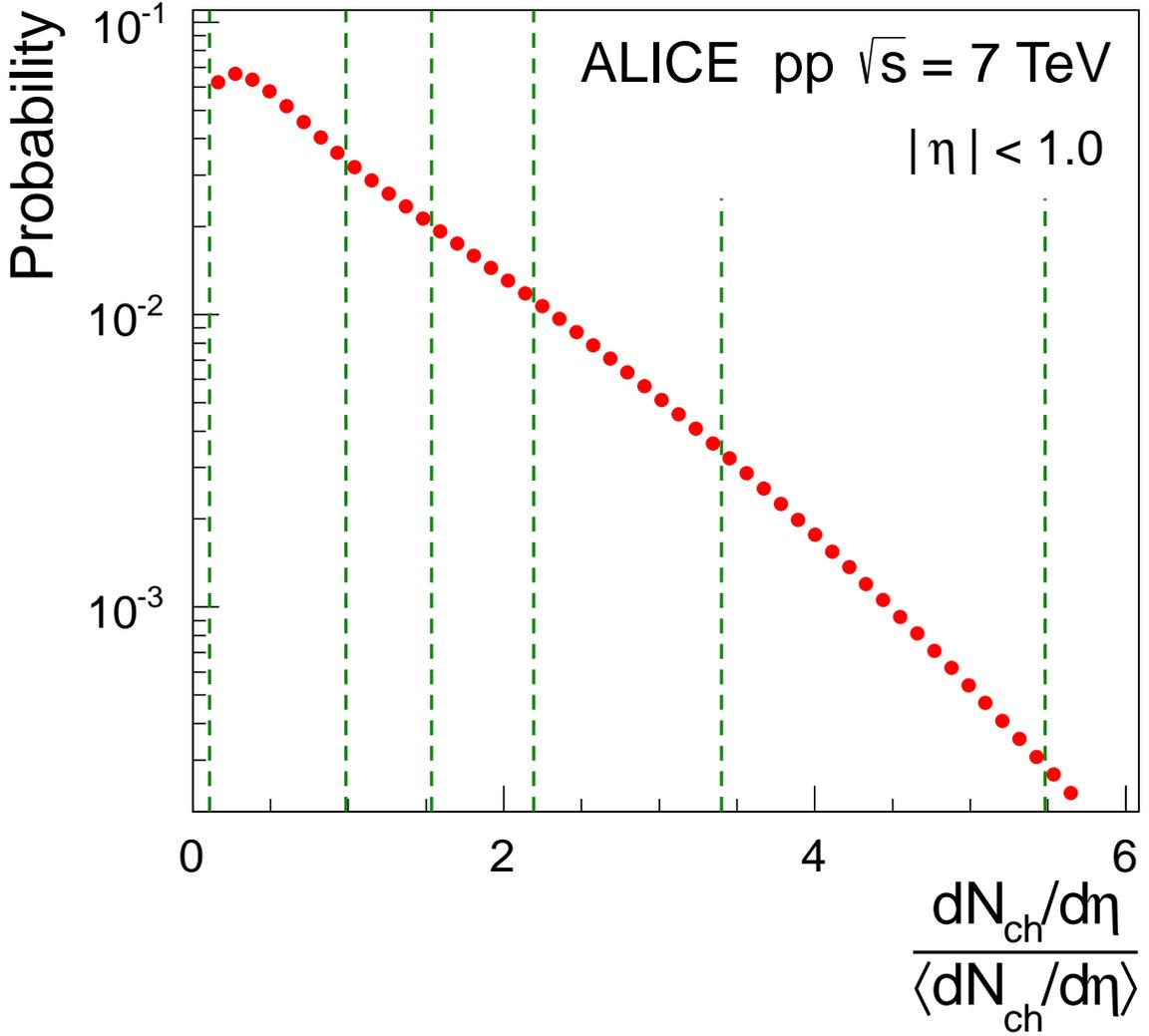}
\caption{\label{fig:mult}
The distribution of the relative charged particle density
$(\dndetach)/\langle\dndetach\rangle$ reconstructed around mid-rapidity
($|\eta| < 1.0$) after correction for SPD inefficiencies.  The
vertical lines indicate the boundaries of the multiplicity intervals
used in this analysis.
}
\end{figure}
%

The charged particle density \dndetach\ is calculated using the number
of tracklets $N_{\rb{trk}}$ reconstructed from hits in the SPD
detector, because the SPD is the only central barrel detector that is
read out for all of the $\mu$-MB trigger.  The tracklets are required
to point to the reconstructed interaction vertex within $\pm 1$~cm in
radial and $\pm 3$~cm in $z$ direction
\cite{Aamodt:2010ft,Aamodt:2010pp}.  Using simulated events, it is
verified that $N_{\rb{trk}}$ is proportional to \dndetach.  For a good
geometrical coverage, only tracklets within $|\eta| < 1$ from events
with $|\vtxz| < 10$~cm are considered.  Since the pseudorapidity
coverage of the SPD changes with the interaction vertex $z$~position
and also with time, due to the varying number of dead channels, a
correction is determined event-by-event from measured data as a
function of \vtxz\ and for each analysed data set separately.
Figure~\ref{fig:mult} shows the resulting distribution of the relative
charged particle density $(\dndetach)/\langle\dndetach\rangle$, where
$\langle\dndetach\rangle = 6.01 \pm
0.01(\textrm{stat.})^{+0.20}_{-0.12}(\textrm{syst.})$ as measured for
inelastic pp collisions with at least one charged particle in $|\eta|
< 1$ \cite{Aamodt:2010pp}.  The use of relative quantities was chosen
in order to facilitate the comparison to other experiments and to
theoretical models, as well as to minimize systematic uncertainties.
The definition of the charged particle multiplicity intervals used in
this analysis is given in \Ta{tab:multbins}, together with the
corresponding mean values of \dndetach.  The present statistics allows
one to cover charged particle densities up to four times the minimum
bias value. 

%
\begin{table*}[t]
\caption{The boundaries of the used charged particle multiplicity
  intervals as defined via the number of SPD tracklets $N_{\rb{trk}}$,
  the corresponding charged particles density ranges and mean values
  $\langle\dndetach(\textrm{bin})\rangle$, as well as the number of
  analyzed minimum bias triggered events in the di-electron
  ($N_{\rb{evt.}}^{\epem}$) and the di-muon channel
  ($N_{\rb{eq. evt.}}^{\mpmm}$).  In the latter case this is the
  equivalent number of events, derived from the number of $\mu$-MB
  triggered events.
}
\begin{tabular}{cccccc}
\label{tab:multbins} 
       Multiplicity interval
     & $N_{\rb{trk}}$ interval
     & \dndetach\ range
     & $\langle\dndetach(\textrm{bin})\rangle$
     & $N_{\rb{evt.}}^{\epem} \times 10^{6}$
     & $N_{\rb{eq. evt.}}^{\mpmm} \times 10^{6}$ \\ \hline
1    & [ 1,  8] &  0.7 --  5.9 &  2.7 & 164.6 & 262.0 \\
2    & [ 9, 13] &  5.9 --  9.2 &  7.1 &  51.1 &  79.5 \\
3    & [14, 19] &  9.2 -- 13.2 & 10.7 &  35.7 &  55.4 \\
4    & [20, 30] & 13.2 -- 20.4 & 15.8 &  28.5 &  44.4 \\
5    & [31, 49] & 20.4 -- 32.9 & 24.1 &   9.7 &  15.3 \\
\end{tabular}
\end{table*}
%

For the \jpsi\ measurement in the di-electron channel tracks are
selected by requiring a minimum \pt\ of 1~\gevc, a pseudorapidity
range of $|\eta| < 0.9$, at least 70 out of possible 159 points
reconstructed in the TPC and an upper limit on the $\chisq/n.d.f.$
from the momentum fit of 2.0.  Furthermore, tracks that are not
pointing back to the primary interaction vertex within 1.0~cm in the
transverse plane and within 3.0~cm in $z$ direction are discarded.  To
further reduce the background from conversion electrons a hit in at
least one of the four innermost ITS layers is also required.  Particle
identification is performed by measuring the specific ionization
\dedx\ in the TPC.  All tracks within $\pm 3 \sigma$ around the
expected \dedx\ signal for electrons and at the same time outside $\pm
3 \sigma$ ($\pm 3.5 \sigma$) around the expectation for protons
(pions) are accepted as electron and positron candidates.  \elp\ and
\elm\ candidates that form a pair with any other candidate with an
invariant mass below 0.1~\gevcc\ are discarded to reduce the amount of
electrons coming from $\gamma$ conversions or \pizero\ Dalitz decays
as well as their contribution to the combinatorial background in the
di-electron invariant mass spectrum.

The invariant mass distributions of the \epem\ pairs are recorded in
intervals of the charged particle multiplicity as measured using the
SPD tracklets.  As an example, the lowest and highest multiplicity
intervals are shown in the two left panels of \Fi{fig:minv}.  The
combinatorial background in each multiplicity interval is well
described by the track rotation method, which consists in rotating one
of the tracks of a \epem\ pair measured in a given event around the
$z$~axis by a random $\phi$-angle in order to remove any correlations.
After subtracting the background, the uncorrected \jpsi\ yields are
obtained by integrating the distribution in the mass range
2.92~--~3.16~\gevcc.  Alternatively, the combinatorial background is
estimated by like-sign distributions, $N^{++} + N^{--}$.  These are
scaled to match the integral of the opposite-sign distributions in the
mass range above the \jpsi\ signal ($3.2 < \minv < 4.9$~\gevcc) in
order to also account for correlated background contributions, which
mainly originates from semi-leptonic charm decays.  Both methods
provide a good description of the combinatorial background and their
comparison is used to evaluate the systematic uncertainty on the
\jpsi\ signal.

For the \jpsi\ analysis in the di-muon channel muon candidates are
selected by requiring that at least one of the two muon candidates
matches a trigger track reconstructed from at least three hits in the
trigger chambers.  This efficiently rejects hadrons produced in the
frontal absorber and then absorbed by the iron wall positioned
in front of the trigger chambers.  Furthermore, a cut $R_{\rb{abs}} >
17.5$~cm is applied, where $R_{\rb{abs}}$ is the radial coordinate of
the track at the end of the frontal absorber ($z = -5.03$~m).  Such a
cut removes muons produced at small angles that have crossed a
significant fraction of the thick beam shield.  Finally, a cut on the
pair rapidity ($2.5 < y < 4$) is applied to reject events very close
to the edge of the spectrometer acceptance.

%
\begin{figure}[t]
\begin{center}
\begin{minipage}[b]{0.49\linewidth}
\includegraphics[width=\linewidth]{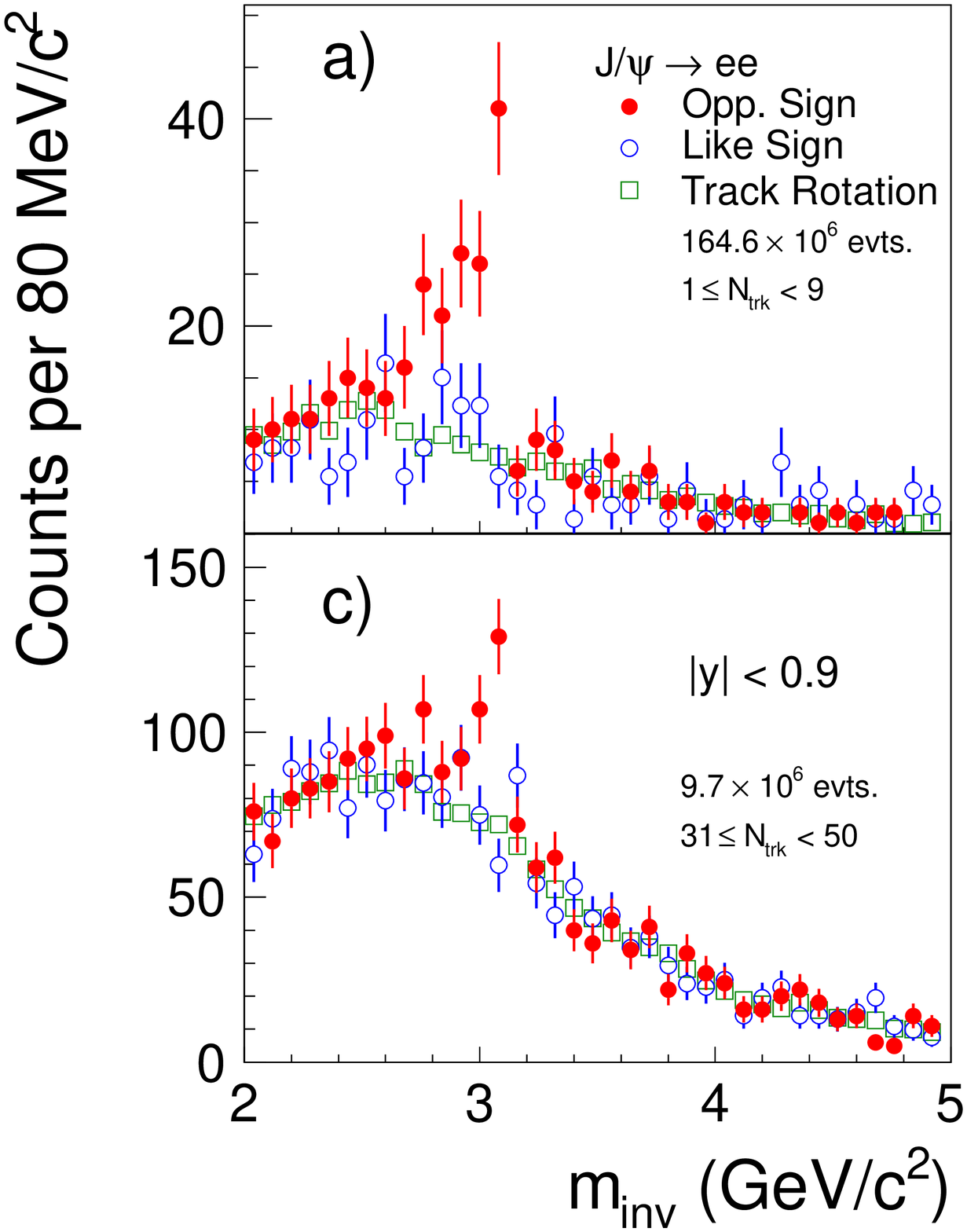}
\end{minipage}
\begin{minipage}[b]{0.49\linewidth}
\includegraphics[width=\linewidth]{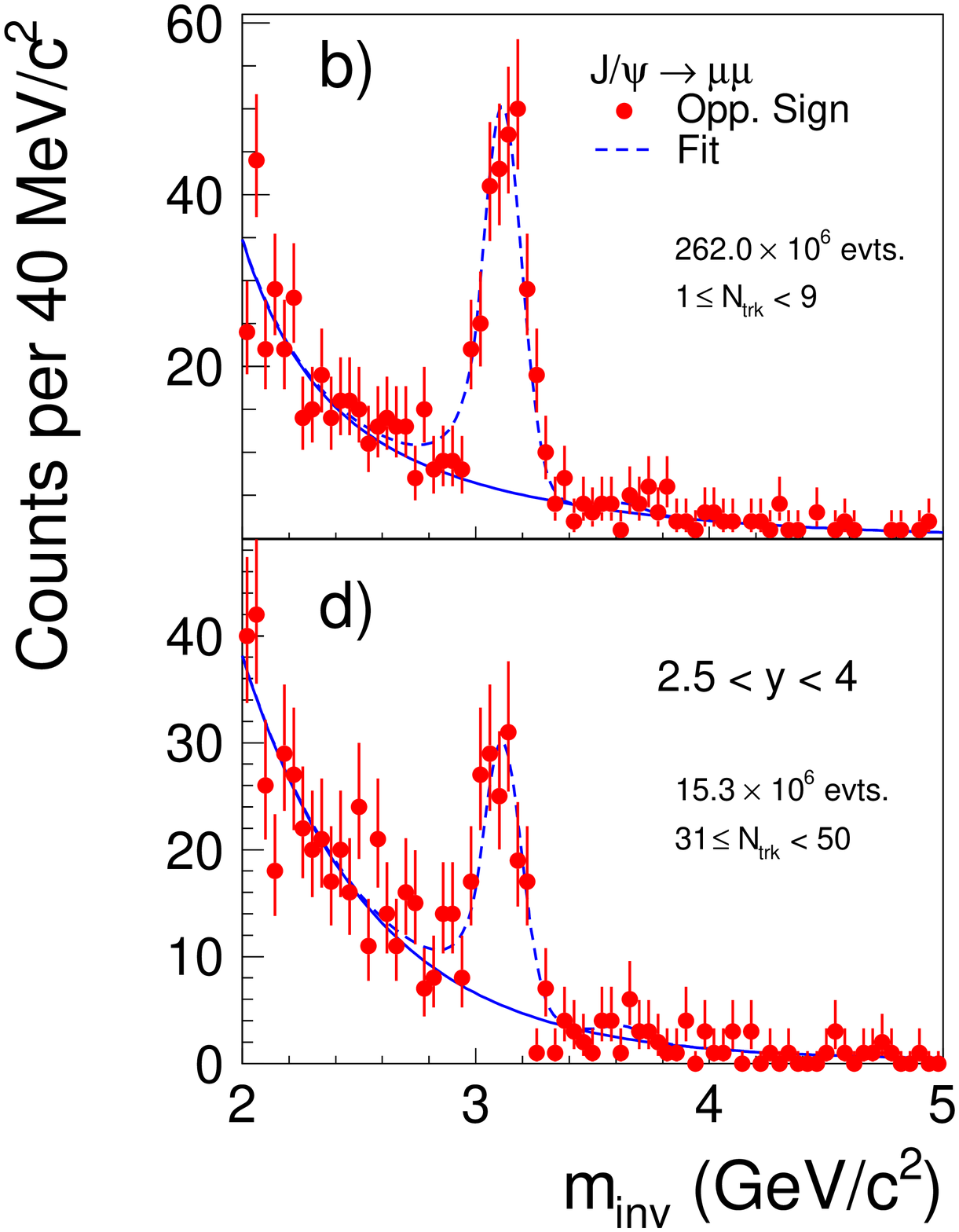}
\end{minipage}
\end{center}
\caption{\label{fig:minv}
Opposite sign invariant mass spectra of the selected electron
[(a) + (c)] and muon [(b) + (d)] pairs (filled symbols) for the
lowest [(a) + (b)] and highest [(c) + (d)] multiplicity intervals.
Also shown are the estimates of the combinatorial background which are
based on a fit to the \mpmm\ pair distributions (solid line), and on
like-sign pairs (open circles), as well as track rotation (open
squares), in the \epem\ case.  The number of events quoted in the
figures refer to the corresponding minimum bias triggered events.
}
\end{figure}
%

%
\begin{figure}[t]
\includegraphics[width=0.85\linewidth]{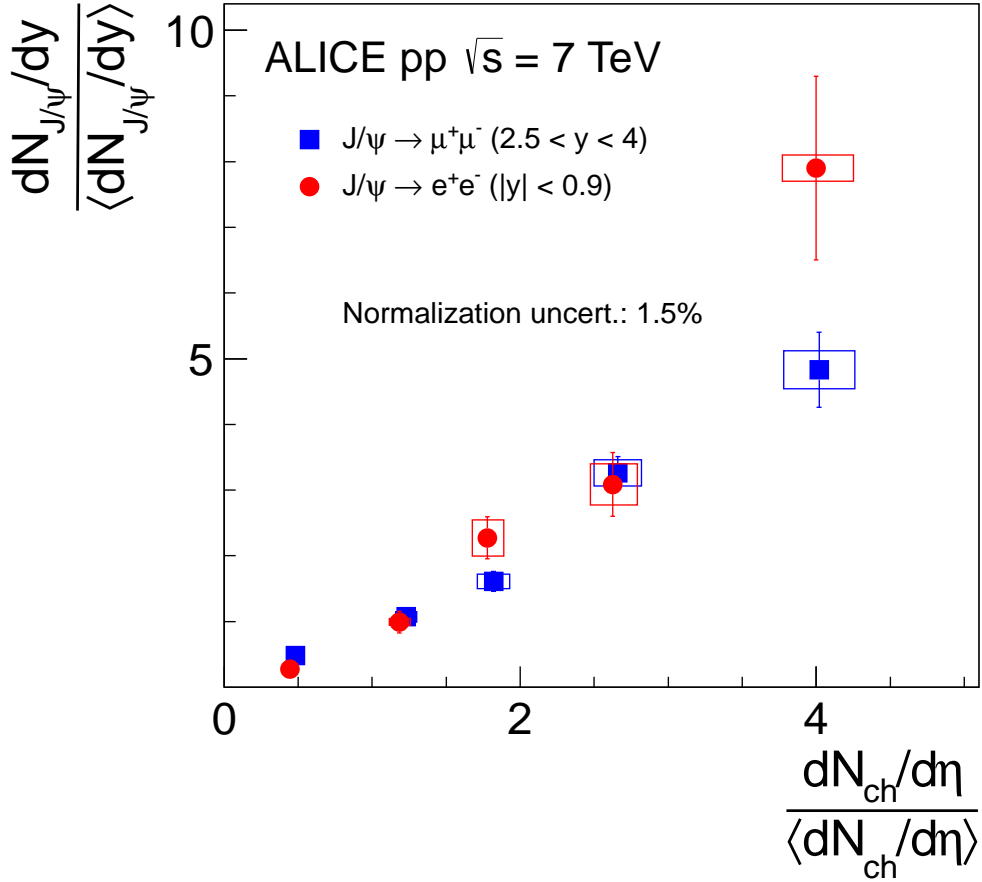}
\caption{\label{fig:reljpsiyield}
\jpsi\ yield \dndyjp\ as a function of the charged particle
multiplicity densities at mid-rapidity \dndetach.  Both values are
normalized by the corresponding value for minimum bias pp collisions
($\langle\dndyjp\rangle$, $\langle\dndetach\rangle$).  Shown are
measurements at forward rapidities ($\jpsi \rightarrow \mpmm$, $2.5 <
y < 4$) and at mid-rapidity ($\jpsi \rightarrow \epem$, $|y| < 0.9$).
The error bars represent the statistical uncertainty on the \jpsi\
yields, while the quadratic sum of the point-by-point systematic
uncertainties on the \jpsi\ yield as well as on \dndetach\ is depicted
as boxes.
}
\end{figure}
%

%
\begin{figure}[t]
\includegraphics[width=0.85\linewidth]{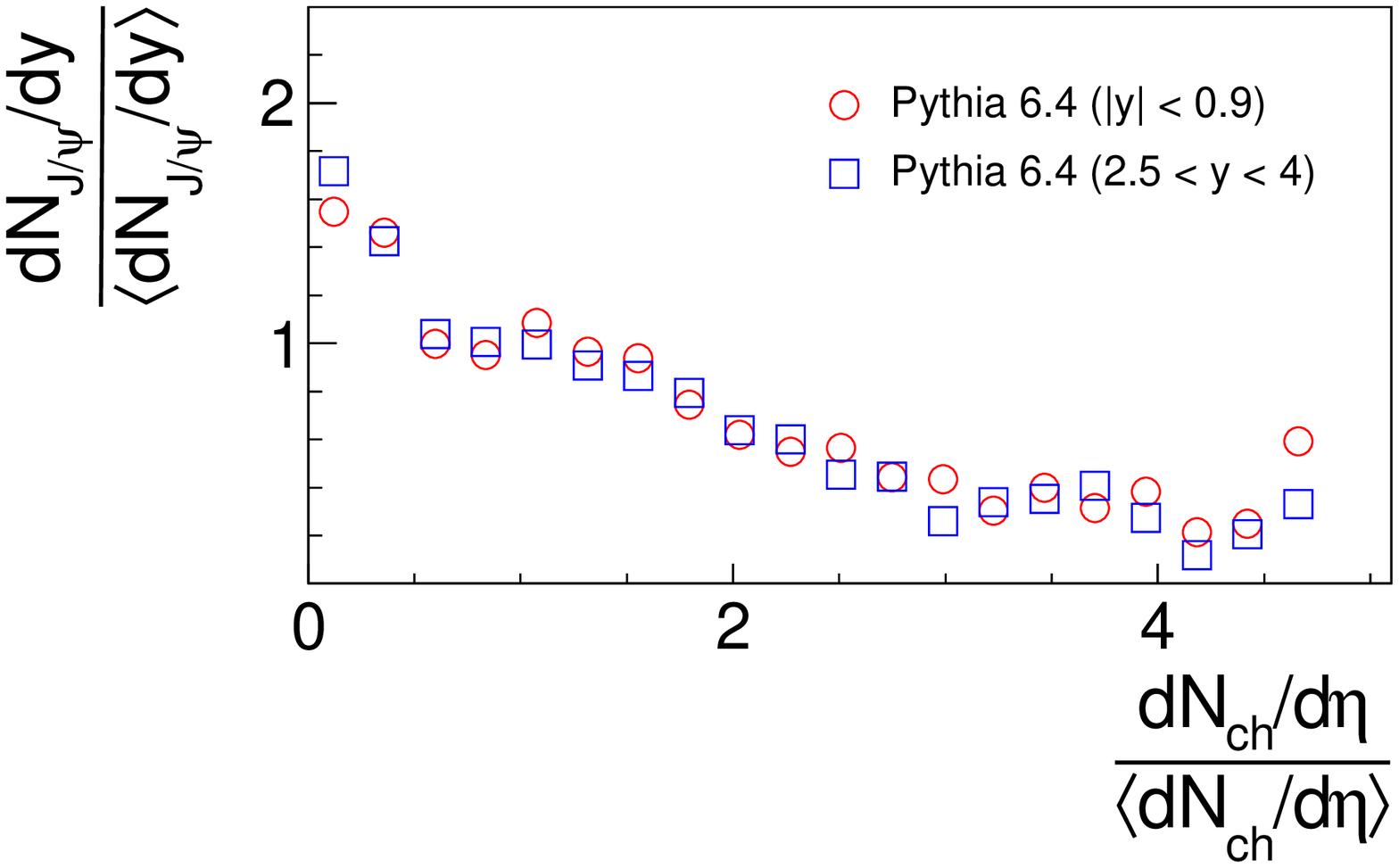}
\caption{\label{fig:pythia}
Relative \jpsi\ yield \dndyjp\ as a function of relative charged
particle multiplicity densities around mid-rapidity \dndetach\ as
calculated with PYTHIA~6.4 in the Perugia~2011 tune 
\cite{Skands:2010ak,Sjostrand:2006za}.  Shown are results for 
directly produced \jpsi\ in hard scatterings via the NRQCD framework
at forward rapidities ($2.5 < y < 4$) and at mid-rapidity ($|y| <
0.9$).
}
\end{figure}
%

The number of \jpsi\ in each multiplicity interval is obtained by
fitting the corresponding di-muon invariant mass distribution in the
range $2 < \minv < 5$~\gevcc.  The line shapes of the \jpsi\ and
$\psi$(2S) are parametrized using Crystal Ball functions
\cite{Gaiser:1982}, while the underlying continuum is fitted with the
sum of two exponential functions.  The parameters of the Crystal Ball
functions are adjusted to the mass distribution of a Monte Carlo
signal sample, obtained by generating \jpsi\ and $\psi$(2S) events
with realistic phase space distributions \cite{Aamodt:2011gj}.  Apart
from the \jpsi\ and $\psi$(2S) signal normalization, only the position
of the \jpsi\ mass pole, as well as its width, are kept as free
parameters in the fit.  Due to the small statistics, the $\psi$(2S)
mass and width are tied to those of the \jpsi, imposing the mass
difference between the two states to be equal to the one given by the
Particle Data Group (PDG) \cite{PDG:2010}, and the ratio of the
resonance widths to be equal to the one obtained by analyzing
reconstructed Monte Carlo events.  Details on the fit technique can be
found in \cite{Aamodt:2011gj}.  The two right panels of \Fi{fig:minv}
show the measured di-muon invariant mass distributions together with
the results of the fit procedure for the lowest and highest
multiplicity intervals.

The results are presented as the ratios of the \jpsi\ yield in a given
multiplicity interval relative to the minimum bias yield.  By
performing simulation studies in intervals of \dndetach\ it was
verified that the geometrical acceptances, as well as the
reconstruction efficiencies, do not depend on \dndetach\ in the range
under consideration here ($\dndetach < 32.9$).  Therefore, these
corrections and their corresponding systematic uncertainties cancel in
the ratio $(\dndyjp)/\langle\dndyjp\rangle$ and only the uncorrected
signal counts have to be divided.  The number of events used for the
normalization of $\langle\dndyjp\rangle$ is corrected for the fraction
of inelastic events not seen by the MB trigger condition.  After
applying acceptance and efficiency corrections these values are in
agreement with those that can be obtained from the numbers quoted in
\cite{Aamodt:2011gj}: $\langle\dndyjp\rangle = (8.2 \pm
0.8(\textrm{stat.}) \pm 1.2 (\textrm{syst.})) \times 10^{-5}$ for
$\jpsi \rightarrow \epem$ in $|y| < 0.9$, and $\langle\dndyjp\rangle
= (5.8 \pm 0.2(\textrm{stat.}) \pm 0.6 (\textrm{syst.})) \times
10^{-5}$ for $\jpsi \rightarrow \mpmm$ in $2.5 < y < 4$.  In the case
of the \jpsi\ yields measured in a given multiplicity interval, no
trigger-related correction is needed, since the trigger efficiency is
100\% for $N_{\rb{trk}} \ge 1$.

The systematic uncertainties are estimated as follows.  In case of the
di-electron analysis, the absolute differences between the resulting
$(\dndyjp)/\langle\dndyjp\rangle$ values obtained by using the
like-sign and the track rotation methods define the uncertainty due to
the background subtraction.  It is found to vary between 2\% and
12\% for the different multiplicity intervals.  For the di-muon
analysis this uncertainty is evaluated by varying the functional form
of the background description (polynomial instead of sum of two
exponential).  It depends on the signal to background ratio and varies
between 3\% and 4\%.  Since for the muon measurement it is not
possible to associate a measured track to the interaction vertex, an
additional systematic uncertainty arises from pile-up events.  Among
the vertices inside these events always the one with the largest
number of associated tracks is chosen as main vertex.  Therefore,
events with very low multiplicities are more likely to have a wrong
assignment and thus this uncertainty is largest in the first
multiplicity interval (6\%), while it is 3\% in the others.  Possible
changes of the \pt~spectra with event multiplicity can introduce a
\dndetach\ dependence of the acceptance and efficiency correction,
thus resulting in an additional systematic uncertainty.  This is
estimated by varying the \ptavg\ of the \jpsi\ spectrum that is used
as input to the determination of the corrections via simulation
between 2.6 and 3.2~\gevc.  A systematic effect of 1.5\% (3.5\%) is
found for the di-electron (di-muon) analysis.  The total systematic
error on $(\dndyjp)/\langle\dndyjp\rangle$ is given by the quadratic
sum of the separated contributions and amounts to 2.5~--12\% depending
on the multiplicity interval for the di-electron result.  In the case
of the di-muon analysis it varies between 8\% in the first and 6\% in
the last multiplicity interval.  An additional global uncertainty of
1.5\% on the normalization of $\langle\dndyjp\rangle$ is introduced by
the correction of the trigger inefficiency for all inelastic
collisions.

The systematic uncertainties on $(\dndetach)/\langle\dndetach\rangle$
are due to deviations from a linear dependence of \dndetach\ on
$N_{\rb{trk}}$ and variations in the $N_{\rb{trk}}$ distributions
which remain after the correction procedure.  The latter are caused by
changes in the SPD acceptance for the different data taking periods.
The first contribution is estimated to be 5\%, while the second is
$\sim 2$\%, as determined by Monte Carlo studies.  In addition, the
systematic uncertainty of the $\langle\dndetach\rangle$ measurement
($^{+3.3 \%}_{-2.0 \%}$) \cite{Aamodt:2010pp} is also included.

%

Figure~\ref{fig:reljpsiyield} shows the relative \jpsi\ yields
measured at forward and at mid-rapidity as a function of the relative
charged particle density around mid-rapidity.  An approximately linear
increase of the relative \jpsi\ yield
$(\dndyjp)/\langle\dndyjp\rangle$ with
$(\dndetach)/\langle\dndetach\rangle$ is observed in both rapidity
ranges.  The enhancement relative to minimum bias \jpsi\ yield is a
factor of approximately 5 at $2.5 < y < 4$ (8 at $|y| < 0.9$) for
events with four times the minimum bias charged particle multiplicity
density.  

A possible interpretation of the observed correlation of the \jpsi\
yield and the charged particle multiplicity might be that \jpsi\
production is always accompanied by a strong hadronic activity, thus
biasing the \dndetach\ distributions to higher values.  Since
this correlation extends over the three units of rapidity between the
mid-rapidity \dndetach\ and the forward rapidity \jpsi\ measurement,
it would have far reaching consequences on any model trying to
describe \jpsi\ production in pp collisions.  

In order to illustrate that the observed behaviour cannot be understood
by a simple $2 \rightarrow 2$ hard partonic scattering scenario, a
prediction by PYTHIA~6.4.25 in the Perugia~2011 tune
\cite{Skands:2010ak,Sjostrand:2006za} is shown in \Fi{fig:pythia} as
an example.  Only \jpsi\ directly produced in hard scatterings via the
NRQCD framework \cite{Bodwin:1994jh} (MSEL=63) are considered, whereas
\jpsi\ resulting from the cluster formation processes are ignored.  A
\jpsi\ cluster is a string formed by a $\ccbar$~pair produced via
parton shower evolution which has an invariant mass that is too low
for the standard Lund string fragmentation procedure and thus does not
correspond to a well defined hard scattering process.  The calculation
shown in \Fi{fig:pythia} is thus the ratio of the multiplicity
distributions generated for minimum bias events and events containing
\jpsi\ from hard scatterings.  It exhibits a decrease of the \jpsi\
multiplicity with respect to the event multiplicity, which indicates
that hard \jpsi\ production, as modeled by PYTHIA~6.4.25, is not
accompanied by an increase of the total hadronic activity.  Further
studies with other models such as PYTHIA~8 \cite{Sjostrand:2007gs} and
Cascade \cite{Jung:2001hx} are needed.  It should be pointed out that
the measurement also includes \jpsi\ from the decay of beauty hadrons,
which is not part of the shown PYTHIA result.  The fraction of \jpsi\
from feed down can change with the event multiplicity and can
therefore contribute to the observed multiplicity dependence.
However, since this contribution is on the order of 10\%
\cite{Aad:2011sp,Aaij:2011jh,Chatrchyan:2011kc} it might be only a
small contribution to the observed differences between model and data.

On the other hand, the increase of the \jpsi\ production with event
multiplicity, as reported here, might be due to MPI.  In this scenario
the multiplicity of charged particles is a direct measurement of the
number of partonic interactions in the pp events.  If the effect of
MPI extends into the regime of hard processes, also the \jpsi\ yield
should scale with the number of partonic collisions and the observed
correlation will result.  It has even been conjectured in
\cite{Strikman:2011ar} that the increase of the \jpsi\ yield with
\dndetach\ and the ridge phenomenon observed in high-multiplicity pp
collisions \cite{Khachatryan:2010gv} could be related.  They might
both be caused by gluon density fluctuations and a special transverse
distribution of the gluon densities inside the nucleon.  The presence
of these fluctuations could significantly increase the probability for
MPI and thus cause the observed rise of the \jpsi\ yield.

The multiplicity dependence measured here will allow a direct
comparison of the \jpsi\ production in pp to the one observed in
heavy-ion collisions.  With a mean value of \dndetach\ of 24.1, the
highest multiplicity interval shown in \Fi{fig:reljpsiyield}, for
instance, corresponds roughly to 45~--~50\% centrality for Cu--Cu
collisions at \sqrtsnn~= 200~GeV \cite{Alver:2010ck}.  In order to
establish whether any evidence for a \jpsi\ suppression is observed
already in pp, a proper normalization is needed.  This could be
provided by a measurement of open charm production in the same
multiplicity bins.  Corresponding studies are currently ongoing.

In summary, \jpsi\ yields are measured for the first time in pp
collisions as a function of the charged particle multiplicity density
\dndetach.  \jpsi\ mesons are detected at mid-rapidity ($|y| < 0.9$)
and forward rapidity ($2.5 < y < 4$), while \dndetach\ is determined
at mid-rapidity ($|\eta| < 1$).  An approximately linear increase of
the \jpsi\ yields with the charged particle multiplicity is observed.
The increase is similar at forward and mid-rapidity, exhibiting an
enhancement relative to minimum bias \jpsi\ yield by a factor of about
5 at $2.5 < y < 4$ (8 at $|y| < 0.9$) for events with four times the
minimum bias charged particle multiplicity.  Our result might either
indicate that \jpsi\ production in pp collisions is always connected
with a strong hadronic activity, or that multi-parton interactions
could also affect the harder momentum scales relevant for quarkonia
production.  Further studies of charged particle multiplicity
dependence of \jpsi, $\Upsilon$, and open charm production, also as a
function of \pt, will shed more light on the nature of the observed
effect.

%
\newenvironment{acknowledgement}{\relax}{\relax}
\begin{acknowledgement}
\section{Acknowledgements}

The ALICE collaboration would like to thank all its engineers and technicians for their invaluable contributions to the construction of the experiment and the CERN accelerator teams for the outstanding performance of the LHC complex.
\\
The ALICE collaboration acknowledges the following funding agencies for their support in building and
running the ALICE detector:
 \\
Calouste Gulbenkian Foundation from Lisbon and Swiss Fonds Kidagan, Armenia;
 \\
Conselho Nacional de Desenvolvimento Cient\'{\i}fico e Tecnol\'{o}gico (CNPq), Financiadora de Estudos e Projetos (FINEP),
Funda\c{c}\~{a}o de Amparo \`{a} Pesquisa do Estado de S\~{a}o Paulo (FAPESP);
 \\
National Natural Science Foundation of China (NSFC), the Chinese Ministry of Education (CMOE)
and the Ministry of Science and Technology of China (MSTC);
 \\
Ministry of Education and Youth of the Czech Republic;
 \\
Danish Natural Science Research Council, the Carlsberg Foundation and the Danish National Research Foundation;
 \\
The European Research Council under the European Community's Seventh Framework Programme;
 \\
Helsinki Institute of Physics and the Academy of Finland;
 \\
French CNRS-IN2P3, the `Region Pays de Loire', `Region Alsace', `Region Auvergne' and CEA, France;
 \\
German BMBF and the Helmholtz Association;
\\
General Secretariat for Research and Technology, Ministry of
Development, Greece;
\\
Hungarian OTKA and National Office for Research and Technology (NKTH);
 \\
Department of Atomic Energy and Department of Science and Technology of the Government of India;
 \\
Istituto Nazionale di Fisica Nucleare (INFN) of Italy;
 \\
MEXT Grant-in-Aid for Specially Promoted Research, Ja\-pan;
 \\
Joint Institute for Nuclear Research, Dubna;
 \\
National Research Foundation of Korea (NRF);
 \\
CONACYT, DGAPA, M\'{e}xico, ALFA-EC and the HELEN Program (High-Energy physics Latin-American--European Network);
 \\
Stichting voor Fundamenteel Onderzoek der Materie (FOM) and the Nederlandse Organisatie voor Wetenschappelijk Onderzoek (NWO), Netherlands;
 \\
Research Council of Norway (NFR);
 \\
Polish Ministry of Science and Higher Education;
 \\
National Authority for Scientific Research - NASR (Autoritatea Na\c{t}ional\u{a} pentru Cercetare \c{S}tiin\c{t}ific\u{a} - ANCS);
 \\
Federal Agency of Science of the Ministry of Education and Science of Russian Federation, International Science and
Technology Center, Russian Academy of Sciences, Russian Federal Agency of Atomic Energy, Russian Federal Agency for Science and Innovations and CERN-INTAS;
 \\
Ministry of Education of Slovakia;
 \\
Department of Science and Technology, South Africa;
 \\
CIEMAT, EELA, Ministerio de Educaci\'{o}n y Ciencia of Spain, Xunta de Galicia (Conseller\'{\i}a de Educaci\'{o}n),
CEA\-DEN, Cubaenerg\'{\i}a, Cuba, and IAEA (International Atomic Energy Agency);
 \\
Swedish Research Council (VR) and Knut $\&$ Alice Wallenberg
Foundation (KAW);
 \\
Ukraine Ministry of Education and Science;
 \\
United Kingdom Science and Technology Facilities Council (STFC);
 \\
The United States Department of Energy, the United States National
Science Foundation, the State of Texas, and the State of Ohio.

\end{acknowledgement}

\newpage

%
\appendix
\section{The ALICE Collaboration}
\label{app:collab}

\begingroup
\small
\begin{flushleft}
B.~Abelev\Irefn{org1234}\And
J.~Adam\Irefn{org1274}\And
D.~Adamov\'{a}\Irefn{org1283}\And
A.M.~Adare\Irefn{org1260}\And
M.M.~Aggarwal\Irefn{org1157}\And
G.~Aglieri~Rinella\Irefn{org1192}\And
A.G.~Agocs\Irefn{org1143}\And
A.~Agostinelli\Irefn{org1132}\And
S.~Aguilar~Salazar\Irefn{org1247}\And
Z.~Ahammed\Irefn{org1225}
% ...
\renewcommand\labelenumi{\textsuperscript{\theenumi}~}
\section*{Affiliation notes}
\renewcommand\theenumi{\roman{enumi}}
\begin{Authlist}
\item \Adef{0}Deceased
\item \Adef{M.V.Lomonosov Moscow State University, D.V.Skobeltsyn Institute of Nuclear Physics, Moscow, Russia}Also at: M.V.Lomonosov Moscow State University, D.V.Skobeltsyn Institute of Nuclear Physics, Moscow, Russia
\item \Adef{Institute of Nuclear Sciences, Belgrade, Serbia}Also at: "Vin\v{c}a" Institute of Nuclear Sciences, Belgrade, Serbia
\end{Authlist}
\section*{Collaboration Institutes}
\renewcommand\theenumi{\arabic{enumi}~}
\begin{Authlist}
\item \Idef{org1279}Benem\'{e}rita Universidad Aut\'{o}noma de Puebla, Puebla, Mexico
\item \Idef{org1220}Bogolyubov Institute for Theoretical Physics, Kiev, Ukraine
\item \Idef{org1262}Budker Institute for Nuclear Physics, Novosibirsk, Russia
\item \Idef{org1292}California Polytechnic State University, San Luis Obispo, California, United States
\item \Idef{org14939}Centre de Calcul de l'IN2P3, Villeurbanne, France
\item \Idef{org1197}Centro de Aplicaciones Tecnol\'{o}gicas y Desarrollo Nuclear (CEADEN), Havana, Cuba
\item \Idef{org1242}Centro de Investigaciones Energ\'{e}ticas Medioambientales y Tecnol\'{o}gicas (CIEMAT), Madrid, Spain
\item \Idef{org1244}Centro de Investigaci\'{o}n y de Estudios Avanzados (CINVESTAV), Mexico City and M\'{e}rida, Mexico
\item \Idef{org1335}Centro Fermi -- Centro Studi e Ricerche e Museo Storico della Fisica ``Enrico Fermi'', Rome, Italy
\item \Idef{org17347}Chicago State University, Chicago, United States
\item \Idef{org1288}Commissariat \`{a} l'Energie Atomique, IRFU, Saclay, France
\item \Idef{org1294}Departamento de F\'{\i}sica de Part\'{\i}culas and IGFAE, Universidad de Santiago de Compostela, Santiago de Compostela, Spain
\item \Idef{org1106}Department of Physics Aligarh Muslim University, Aligarh, India
\item \Idef{org1121}Department of Physics and Technology, University of Bergen, Bergen, Norway
\item \Idef{org1162}Department of Physics, Ohio State University, Columbus, Ohio, United States
\item \Idef{org1300}Department of Physics, Sejong University, Seoul, South Korea
\item \Idef{org1268}Department of Physics, University of Oslo, Oslo, Norway
\item \Idef{org1145}Dipartimento di Fisica dell'Universit\`{a} and Sezione INFN, Cagliari, Italy
\item \Idef{org1270}Dipartimento di Fisica dell'Universit\`{a} and Sezione INFN, Padova, Italy
\item \Idef{org1315}Dipartimento di Fisica dell'Universit\`{a} and Sezione INFN, Trieste, Italy
\item \Idef{org1132}Dipartimento di Fisica dell'Universit\`{a} and Sezione INFN, Bologna, Italy
\item \Idef{org1285}Dipartimento di Fisica dell'Universit\`{a} `La Sapienza' and Sezione INFN, Rome, Italy
\item \Idef{org1154}Dipartimento di Fisica e Astronomia dell'Universit\`{a} and Sezione INFN, Catania, Italy
\item \Idef{org1290}Dipartimento di Fisica `E.R.~Caianiello' dell'Universit\`{a} and Gruppo Collegato INFN, Salerno, Italy
\item \Idef{org1312}Dipartimento di Fisica Sperimentale dell'Universit\`{a} and Sezione INFN, Turin, Italy
\item \Idef{org1103}Dipartimento di Scienze e Tecnologie Avanzate dell'Universit\`{a} del Piemonte Orientale and Gruppo Collegato INFN, Alessandria, Italy
\item \Idef{org1114}Dipartimento Interateneo di Fisica `M.~Merlin' and Sezione INFN, Bari, Italy
\item \Idef{org1237}Division of Experimental High Energy Physics, University of Lund, Lund, Sweden
\item \Idef{org1192}European Organization for Nuclear Research (CERN), Geneva, Switzerland
\item \Idef{org1227}Fachhochschule K\"{o}ln, K\"{o}ln, Germany
\item \Idef{org1122}Faculty of Engineering, Bergen University College, Bergen, Norway
\item \Idef{org1136}Faculty of Mathematics, Physics and Informatics, Comenius University, Bratislava, Slovakia
\item \Idef{org1274}Faculty of Nuclear Sciences and Physical Engineering, Czech Technical University in Prague, Prague, Czech Republic
\item \Idef{org1229}Faculty of Science, P.J.~\v{S}af\'{a}rik University, Ko\v{s}ice, Slovakia
\item \Idef{org1184}Frankfurt Institute for Advanced Studies, Johann Wolfgang Goethe-Universit\"{a}t Frankfurt, Frankfurt, Germany
\item \Idef{org1215}Gangneung-Wonju National University, Gangneung, South Korea
\item \Idef{org1212}Helsinki Institute of Physics (HIP) and University of Jyv\"{a}skyl\"{a}, Jyv\"{a}skyl\"{a}, Finland
\item \Idef{org1203}Hiroshima University, Hiroshima, Japan
\item \Idef{org1329}Hua-Zhong Normal University, Wuhan, China
\item \Idef{org1254}Indian Institute of Technology, Mumbai, India
\item \Idef{org36378}Indian Institute of Technology Indore (IIT), Indore, India
\item \Idef{org1266}Institut de Physique Nucl\'{e}aire d'Orsay (IPNO), Universit\'{e} Paris-Sud, CNRS-IN2P3, Orsay, France
\item \Idef{org1277}Institute for High Energy Physics, Protvino, Russia
\item \Idef{org1249}Institute for Nuclear Research, Academy of Sciences, Moscow, Russia
\item \Idef{org1320}Nikhef, National Institute for Subatomic Physics and Institute for Subatomic Physics of Utrecht University, Utrecht, Netherlands
\item \Idef{org1250}Institute for Theoretical and Experimental Physics, Moscow, Russia
\item \Idef{org1230}Institute of Experimental Physics, Slovak Academy of Sciences, Ko\v{s}ice, Slovakia
\item \Idef{org1127}Institute of Physics, Bhubaneswar, India
\item \Idef{org1275}Institute of Physics, Academy of Sciences of the Czech Republic, Prague, Czech Republic
\item \Idef{org1139}Institute of Space Sciences (ISS), Bucharest, Romania
\item \Idef{org27399}Institut f\"{u}r Informatik, Johann Wolfgang Goethe-Universit\"{a}t Frankfurt, Frankfurt, Germany
\item \Idef{org1185}Institut f\"{u}r Kernphysik, Johann Wolfgang Goethe-Universit\"{a}t Frankfurt, Frankfurt, Germany
\item \Idef{org1177}Institut f\"{u}r Kernphysik, Technische Universit\"{a}t Darmstadt, Darmstadt, Germany
\item \Idef{org1256}Institut f\"{u}r Kernphysik, Westf\"{a}lische Wilhelms-Universit\"{a}t M\"{u}nster, M\"{u}nster, Germany
\item \Idef{org1246}Instituto de Ciencias Nucleares, Universidad Nacional Aut\'{o}noma de M\'{e}xico, Mexico City, Mexico
\item \Idef{org1247}Instituto de F\'{\i}sica, Universidad Nacional Aut\'{o}noma de M\'{e}xico, Mexico City, Mexico
\item \Idef{org23333}Institut of Theoretical Physics, University of Wroclaw
\item \Idef{org1308}Institut Pluridisciplinaire Hubert Curien (IPHC), Universit\'{e} de Strasbourg, CNRS-IN2P3, Strasbourg, France
\item \Idef{org1182}Joint Institute for Nuclear Research (JINR), Dubna, Russia
\item \Idef{org1143}KFKI Research Institute for Particle and Nuclear Physics, Hungarian Academy of Sciences, Budapest, Hungary
\item \Idef{org1199}Kirchhoff-Institut f\"{u}r Physik, Ruprecht-Karls-Universit\"{a}t Heidelberg, Heidelberg, Germany
\item \Idef{org20954}Korea Institute of Science and Technology Information
\item \Idef{org1160}Laboratoire de Physique Corpusculaire (LPC), Clermont Universit\'{e}, Universit\'{e} Blaise Pascal, CNRS--IN2P3, Clermont-Ferrand, France
\item \Idef{org1194}Laboratoire de Physique Subatomique et de Cosmologie (LPSC), Universit\'{e} Joseph Fourier, CNRS-IN2P3, Institut Polytechnique de Grenoble, Grenoble, France
\item \Idef{org1187}Laboratori Nazionali di Frascati, INFN, Frascati, Italy
\item \Idef{org1232}Laboratori Nazionali di Legnaro, INFN, Legnaro, Italy
\item \Idef{org1125}Lawrence Berkeley National Laboratory, Berkeley, California, United States
\item \Idef{org1234}Lawrence Livermore National Laboratory, Livermore, California, United States
\item \Idef{org1251}Moscow Engineering Physics Institute, Moscow, Russia
\item \Idef{org1140}National Institute for Physics and Nuclear Engineering, Bucharest, Romania
\item \Idef{org1165}Niels Bohr Institute, University of Copenhagen, Copenhagen, Denmark
\item \Idef{org1109}Nikhef, National Institute for Subatomic Physics, Amsterdam, Netherlands
\item \Idef{org1283}Nuclear Physics Institute, Academy of Sciences of the Czech Republic, \v{R}e\v{z} u Prahy, Czech Republic
\item \Idef{org1264}Oak Ridge National Laboratory, Oak Ridge, Tennessee, United States
\item \Idef{org1189}Petersburg Nuclear Physics Institute, Gatchina, Russia
\item \Idef{org1170}Physics Department, Creighton University, Omaha, Nebraska, United States
\item \Idef{org1157}Physics Department, Panjab University, Chandigarh, India
\item \Idef{org1112}Physics Department, University of Athens, Athens, Greece
\item \Idef{org1152}Physics Department, University of Cape Town, iThemba LABS, Cape Town, South Africa
\item \Idef{org1209}Physics Department, University of Jammu, Jammu, India
\item \Idef{org1207}Physics Department, University of Rajasthan, Jaipur, India
\item \Idef{org1200}Physikalisches Institut, Ruprecht-Karls-Universit\"{a}t Heidelberg, Heidelberg, Germany
\item \Idef{org1325}Purdue University, West Lafayette, Indiana, United States
\item \Idef{org1281}Pusan National University, Pusan, South Korea
\item \Idef{org1176}Research Division and ExtreMe Matter Institute EMMI, GSI Helmholtzzentrum f\"ur Schwerionenforschung, Darmstadt, Germany
\item \Idef{org1334}Rudjer Bo\v{s}kovi\'{c} Institute, Zagreb, Croatia
\item \Idef{org1298}Russian Federal Nuclear Center (VNIIEF), Sarov, Russia
\item \Idef{org1252}Russian Research Centre Kurchatov Institute, Moscow, Russia
\item \Idef{org1224}Saha Institute of Nuclear Physics, Kolkata, India
\item \Idef{org1130}School of Physics and Astronomy, University of Birmingham, Birmingham, United Kingdom
\item \Idef{org1338}Secci\'{o}n F\'{\i}sica, Departamento de Ciencias, Pontificia Universidad Cat\'{o}lica del Per\'{u}, Lima, Peru
\item \Idef{org1316}Sezione INFN, Trieste, Italy
\item \Idef{org1271}Sezione INFN, Padova, Italy
\item \Idef{org1313}Sezione INFN, Turin, Italy
\item \Idef{org1286}Sezione INFN, Rome, Italy
\item \Idef{org1146}Sezione INFN, Cagliari, Italy
\item \Idef{org1133}Sezione INFN, Bologna, Italy
\item \Idef{org1115}Sezione INFN, Bari, Italy
\item \Idef{org1155}Sezione INFN, Catania, Italy
\item \Idef{org1322}Soltan Institute for Nuclear Studies, Warsaw, Poland
\item \Idef{org36377}Nuclear Physics Group, STFC Daresbury Laboratory, Daresbury, United Kingdom
\item \Idef{org1258}SUBATECH, Ecole des Mines de Nantes, Universit\'{e} de Nantes, CNRS-IN2P3, Nantes, France
\item \Idef{org1304}Technical University of Split FESB, Split, Croatia
\item \Idef{org1168}The Henryk Niewodniczanski Institute of Nuclear Physics, Polish Academy of Sciences, Cracow, Poland
\item \Idef{org17361}The University of Texas at Austin, Physics Department, Austin, TX, United States
\item \Idef{org1173}Universidad Aut\'{o}noma de Sinaloa, Culiac\'{a}n, Mexico
\item \Idef{org1296}Universidade de S\~{a}o Paulo (USP), S\~{a}o Paulo, Brazil
\item \Idef{org1149}Universidade Estadual de Campinas (UNICAMP), Campinas, Brazil
\item \Idef{org1239}Universit\'{e} de Lyon, Universit\'{e} Lyon 1, CNRS/IN2P3, IPN-Lyon, Villeurbanne, France
\item \Idef{org1205}University of Houston, Houston, Texas, United States
\item \Idef{org20371}University of Technology and Austrian Academy of Sciences, Vienna, Austria
\item \Idef{org1222}University of Tennessee, Knoxville, Tennessee, United States
\item \Idef{org1310}University of Tokyo, Tokyo, Japan
\item \Idef{org1318}University of Tsukuba, Tsukuba, Japan
\item \Idef{org21360}Eberhard Karls Universit\"{a}t T\"{u}bingen, T\"{u}bingen, Germany
\item \Idef{org1225}Variable Energy Cyclotron Centre, Kolkata, India
\item \Idef{org1306}V.~Fock Institute for Physics, St. Petersburg State University, St. Petersburg, Russia
\item \Idef{org1323}Warsaw University of Technology, Warsaw, Poland
\item \Idef{org1179}Wayne State University, Detroit, Michigan, United States
\item \Idef{org1260}Yale University, New Haven, Connecticut, United States
\item \Idef{org1332}Yerevan Physics Institute, Yerevan, Armenia
\item \Idef{org15649}Yildiz Technical University, Istanbul, Turkey
\item \Idef{org1301}Yonsei University, Seoul, South Korea
\item \Idef{org1327}Zentrum f\"{u}r Technologietransfer und Telekommunikation (ZTT), Fachhochschule Worms, Worms, Germany
\end{Authlist}
\endgroup

%

%
\end{document}